\documentstyle[aps,epsf,psfig]{revtex}

\newcommand{\be}{\begin{equation}}
\newcommand{\ee}{\end{equation}}
\newcommand{\bea}{\begin{eqnarray}}
\newcommand{\eea}{\end{eqnarray}}

\newcommand{\shalf}{\mbox{$\textstyle \frac{1}{\sqrt{2}}$}}
\newcommand{\ket}[1]{ | \, #1  \rangle}

\begin{document}
\draft\onecolumn
\title{Realistic teleportation with linear optical elements}
\author{Christian~Trump, Dagmar~Bru\ss\ and Maciej~Lewenstein }
\address{Inst. f\"{u}r Theoret. Physik, Universit\"{a}t Hannover, 
Appelstr. 2, D-30167 Hannover, Germany
}
\date{Received \today}
\maketitle
\begin{abstract}
We calculate the highest possible information gain in a  measurement
of entangled states when employing
a beamsplitter. The result is used to evaluate the
 fidelity, averaged over all unknown inputs,
 in a realistic teleportation protocol that takes account of the imperfect
 detection of  Bell states. Finally, we introduce a probabilistic
 teleportation scheme, where measurements are made in a partially
 entangled basis. 
\end{abstract}
In the recent years some striking features of
quantum entanglement have been unveiled in the context of
quantum information \cite{tmr}.
One of them  is the possibility of teleporting an unknown state 
between two distant  locations \cite{tele}.
Teleportation has been realised experimentally for qubits
\cite{zeilinger,demartini}. A fundamental ingredient
in the teleportation protocol is a Bell measurement, i.e. a projection
onto the four Bell states
\bea
\ket{\psi^\pm}&=&\shalf (\ket{01}\pm\ket{10})\ , \nonumber \\
\ket{\phi^\pm}&=&\shalf (\ket{00}\pm\ket{11})\ .
\label{bell}
\eea
In this paper we study teleportation and Bell measurements with
realistic resources.
The teleportation channel, which consists ideally of a maximally
entangled state as in (\ref{bell}), can nowadays be readily 
provided experimentally by entangled photons
\cite{tittel,kwiat,bouwmeester}. 
Implementing  teleportation with photons means having to 
perform a Bell measurement, i.e. a joint measurement, on
two photons. 
Using non-linear devices this would be in principle easy to do, but 
the efficiency of present day non-linear optical elements is not
sufficient to implement Bell measurements. 
Standard optical tools, like e.g. a 
beamsplitter, act in a linear way on photons. 
\par 
It has been shown in \cite{norbert} that it is impossible to distinguish
the four Bell states unambigously with linear optical elements.
In the case of a beamsplitter with 50\% transmission this is 
easy to understand, as  both  $\ket{\phi^+}$ and 
$\ket{\phi^-}$ as input state will lead to
 two identically polarised photons 
 in one of the outgoing paths.
The topics we address in this letter are outlined as follows: 
First, we ask what is the highest
information one can retrieve from a Bell measurement with a linear optical
element, namely a beamsplitter?
This question has also been discussed recently in 
\cite{norbert2}, for a maximally entangled basis.
We will in addition
 consider the case of measurements in a partially entangled
basis, which can be more  successful than in the case of maximal
 entanglement, depending on the parameters of the beamsplitter.
We will then explain a strategy for teleportation,
 using such imperfect measurements, and
calculate the best teleportation fidelity. In the case of 
measurements in a partially entangled basis
teleportation can be implemented probabilistically.
Note that our approach uses a maximally entangled state shared between
Alice and Bob,
and is therefore  different from the scheme considered
 in \cite{mor}. 
 \par
Let us first study the gain of information in a Bell measurement
using a beamsplitter. Our notation is the same as in \cite{norbert},
and we also use their method to represent
the Bell states in matrix form.
We denote the photon modes with the 
creation operators $a^\dagger_{H,V}$ and  $b^\dagger_{H,V}$, where $a$ and
$b$ stand for  two different directions of propagation,
and the indices $H$ and $V$ denote horizontal and vertical polarisation.
Thus, the  Bell states are 
\bea
\ket{\psi^\pm}&=&\shalf (a_H^\dagger b_V^\dagger\pm a_V^\dagger b_H^\dagger)\ket{0}
  \ , \nonumber \\
\ket{\phi^\pm}&=&\shalf (a_H^\dagger b_H^\dagger\pm a_V^\dagger b_V^\dagger)
\ket{0}\ .
\label{bell2}
\eea
The action of a  realistic beamsplitter is given by  \cite{walls} 
\be
\left( \begin{array}{c} 
               {a_H^\dagger}' \\ {a_V^\dagger}'\\ 
               {b_H^\dagger}' \\ {b_V^\dagger}' \end{array}
\right) = \left( 
\begin{array}{cccc} \cos{\theta_H}e^{i \varphi_H}&0&
              \sin{\theta_H}e^{i \chi_H}& 0\\
              0&\cos{\theta_V}e^{i \varphi_V}&0&
              \sin{\theta_V}e^{i \chi_V}\\
              -\sin{\theta_H}&0&
              \cos{\theta_H}& 0\\
              0& -\sin{\theta_V}&0&
              \cos{\theta_V}\\
                \end{array}
\right) \left( \begin{array}{c} 
               {a_H^\dagger} \\ {a_V^\dagger}\\ 
               {b_H^\dagger} \\ {b_V^\dagger} \end{array}\right)
               \ .
\ee
Here $\cos{\theta_{H(V)}}$ is the transmission coefficient for horizontal
(vertical) polarisation, and the phases of the transmitted (reflected) beam
are given by $\varphi \, (\chi)$. These parameters can be set by the
experimentalist.
\par 
The gain of information when making a Bell measurement 
is given, due to Bayes theorem,  as follows:
\be
\Delta S = S_i -S_f = -\sum_i p_i \log{p_i}+\sum_{i,j}p(\psi_i,\kappa_j)
\log{p(\psi_i|\kappa_j)} \ .
\label{info}
\ee 
All logarithms in this paper are taken to base 2.
Here $\psi_i$ enumerates the four Bell states, and $\kappa_j$ denotes
the 10 possible different product
states in the two-photon manifold,
 for example $a_H^\dagger b_V^\dagger \ket{0}$. These product 
states can in principle be distinguished unambiguously. 
 In our calculations
the initial probabilities $p_i$ for the four Bell states are 
taken to be identical.
\par Calculating $\Delta S$ numerically as a function of  
 $\theta_{H}$ and $\theta_V$ leads to the results presented in Fig.
\ref{fig:bell}. The highest possible information gain is 
$\Delta S_{max} =1.5$, which is
reached by a  beamsplitter with 50\% transmission 
for both polarizations.
\par
Which setting of the beamsplitter leads to the highest gain of
information for less than maximally entangled states?
We denote a basis of  partially entangled states as
\bea
\ket{\psi^-}'&=&(x a_H^\dagger b_V^\dagger - 
   \sqrt{1-x^2}a_V^\dagger b_H^\dagger)\ket{0}\ , \ \ \
   \ket{\psi^+}'=(\sqrt{1-x^2} a_H^\dagger b_V^\dagger + 
  x a_V^\dagger b_H^\dagger)\ket{0}
  \ , \nonumber \\
\ket{\phi^-}'&=&(x a_H^\dagger b_H^\dagger - 
   \sqrt{1-x^2}a_V^\dagger b_V^\dagger)\ket{0}\ , \ \ \
   \ket{\phi^+}'=(\sqrt{1-x^2} a_H^\dagger b_H^\dagger + 
  x a_V^\dagger b_V^\dagger)\ket{0}
\ .
\label{part}
\eea
Repeating the calculation from above one now finds the 
information gain as a function of the degree of entanglement $x$.
An example for $x^2=0.1$ is shown in Fig. \ref{fig:part}.
Here the {\it minima} correspond to equal reflection and transmission.
The maximum information gain is higher than for Bell states,
and will reach $\Delta S_{max}=2$ for $x=0$.
In the example $x^2=0.1$ we find $\Delta S_{max}=1.52$.
\par 
How can one use such  imperfect measurements for 
realistic teleportation,
and what is the fidelity one can achieve?
In order to answer this question one has to develop a strategy that 
copes with the cases where the Bell states have not been detected
unambiguously. 
As usually in teleportation, Alice and Bob share a maximally entangled
state, and Alice wants to teleport an unknown state $\ket{\tau}$ to Bob.
The protocol we suggest progresses as follows:
\begin{itemize}
\item[-] Alice fixes the parameters for her beamsplitter and prepares
a look-up table that tells her the conditional probabilities
$p(\psi_i|\kappa_j)$.
\item[-] Alice does an imperfect ``Bell measurement'' on the unknown state 
$\ket{\tau}$ and
her part of the entangled state, and finds a detector coincidence 
$\kappa_j$.
\item[-] Alice uses her look-up table to determine for which 
$i$ the conditional probability $p(\psi_i|\kappa_j)$ is 
maximal, and informs Bob via classical communication about the result.
\item[-] Bob rotates his state according to Alices most likely
Bell state $\psi_i$. 
\end{itemize}
With this protocol different input states will be teleported with different
fidelity. If one does not want to restrict the set of possible inputs,
a meaningful figure of merit will be an averaged fidelity, namely
\be
\bar{F}= \int_\tau d\tau F(\ket{\tau})\ ,
\ee
where $d\tau$ symbolizes an appropriate integration measure for
the input state. In this paper, the
fidelity  is weighted with a constant probability
for  each input state.
\par 
Performing the numerics for  this strategy leads to the results given
in Fig. \ref{fig:fidel}: the maximal averaged fidelity is again reached for
a  beamsplitter
 with 50\% transmission, and is $\bar{F}_{max}\approx 0.88$.
Remember that the value for the fidelity for ``teleportation'' without using 
entanglement, i.e. a projection measurement by Alice and 
subsequent state 
preparation by Bob, is $F_{class}=2/3$. Therefore, even with
one linear optical element it is in principle possible to 
demonstrate teleportation of an unknown quantum state
in an experiment that takes all Bell states  into account.
 Using our protocol one can, in principle,
arrive at higher fidelities than in the existing experiments,
which reach a fidelity of $F=0.82\pm 0.01$ \cite{dik},
with an efficiency of 25\%.
\par 
In the remainder of this paper we will discuss teleportation with
measurements in a partially entangled basis. 
This kind of teleportation works probabilistically, i.e. 
one succeeds with
a certain probability  and then knows to have succeeded.
The unsuccessful runs have to be discarded.
To our knowledge, this
idea has not yet been studied elsewhere. It is motivated by the fact that 
partially entangled states are easier to distinguish than 
maximally entangled states, as shown in Figs. \ref{fig:bell}
and \ref{fig:part}. 
\par
In order to present the idea of probabilistic teleportation,
for simplicity we will go back to the following notation of 
a basis of partially entangled states:
\bea
\ket{\psi^-}'&=&x\ket{01}-\sqrt{1-x^2}\ket{10}\ , \ \ \ 
\ket{\psi^+}'=\sqrt{1-x^2}\ket{01}+x\ket{10} \ ,\nonumber \\
\ket{\phi^-}'&=&x\ket{00}-\sqrt{1-x^2}\ket{11}\ , \ \ \ 
\ket{\phi^+}'=\sqrt{1-x^2}\ket{00}+x\ket{11}\ .
\label{part2}
\eea
In this basis we can rewrite the total state before
Alice's measurement as
\bea
(\alpha\ket{0}+\beta\ket{1})\otimes\shalf(\ket{01}+\ket{10})&=&
\frac{1}{\sqrt{2}xy}\left[ 
\ket{\psi^+}'(\alpha x\ket{0}+\beta y\ket{1})+
\ket{\psi^-}'(\alpha y\ket{0}-x\beta\ket{1}) \right.\nonumber \\
&& \hspace{1.cm} \left. +\ket{\phi^+}'(\beta x\ket{0}+\alpha y\ket{1})+
\ket{\phi^-}'(\beta y\ket{0}-x\alpha\ket{1})
\right] \ ,
\eea
where we have introduced $y=\sqrt{1-x^2}$.
Without loss of generality we assume $x\geq y$.
After Alice finds one of the four basis states (here we assume
the detection to be perfect), she tells Bob the result, and
he performs a positive operator measurement (POVM).
 Let us consider the case where Alice finds $\ket{\psi^+}'$,
the other cases work in an analogous way. For this case the POVM elements
are given by
\be
A_1=\left( \begin{array}{cc} \frac{y}{x} & 0\\
              0 & 1  \end{array}\right) \ , \ \  \
 A_2=\left( \begin{array}{cc} 1-\frac{y}{x} & 0\\
              0 & 0  \end{array}\right) \ .             
\ee
With probability $p=1/(2x^2)$  this procedure will lead to
the correct state on Bob's side.
\par 
At the end of this article we formulate the  problem of
choosing the best degree of entanglement for the basis of
Alice's measurement: there will be
a trade-off between the information gain (which is higher for
less entangled states) and the probability of successful 
teleportation (which is higher for more entangled states).  
It remains an open question which degree of entanglement one
has to choose in a realistic experiment in order to achieve the
highest teleportation fidelity.
 \par
 In conclusion, we have found that a  beamsplitter 
 with 50\% transmission leads to
 the highest information gain in a Bell measurement, and discussed
 the optimal parameters of a beamsplitter for measuring 
 partially entangled states. 
 We have then presented a strategy for realistic teleportation
 with  imperfect Bell measurements, using
  the ``most likely''  Bell state.
 The  fidelity, averaged over all possible 
 inputs to be teleported, was found
 to be $\bar{F}_{max}\approx 0.88$ 
  with our  protocol. Finally, we introduced the
  concept of probabilistic teleportation, using measurements in
  a partially entangled basis, with a subsequent POVM on Bob's side.  
\par
We would like to thank John Calsamiglia, Norbert L\"utkenhaus and
Harald Weinfurter for helpful discussions. We acknowledge 
support  by DFG Schwerpunkt ``Quanteninformationsverarbeitung'',
and by the European IST Programme EQUIP.

\newpage

\begin{figure}[hbt]
\vspace*{+2cm}
\vspace*{1cm}
\caption{Information gain for Bell states
as a function of the transmission coefficient for
vertical and horizontal polarization. The maxima correspond to 
transmission equals reflection for both polarizations.
}
\label{fig:bell}
\end{figure}

\begin{figure}[hbt]
\vspace*{+2cm}
\vspace*{1cm}
\caption{Information gain for partially entangled states with $x^2=0.1$
as a function of the transmission coefficient for
vertical and horizontal polarization. The minima correspond to 
transmission equals reflection for both polarizations.
}
\label{fig:part}
\end{figure}

\begin{figure}[hbt]
\vspace*{+2cm}
\vspace*{1cm}
\caption{Averaged fidelity for teleportation of an unknown state,
as a function of the transmission coefficient for
vertical and horizontal polarization.
}
\label{fig:fidel}
\end{figure}

\end{document}